\documentclass[aps,prd,showpacs,twocolumn,floatfix]{revtex4}
\usepackage{graphicx}
\usepackage{amsfonts}
\usepackage{amssymb}
\usepackage{amsmath}
\usepackage{flafter}
\usepackage{epstopdf}

\begin{document}

\title{Bose and Fermi gases in the early universe with self-gravitational effect}

\newcommand*{\PKU}{School of Physics and State Key Laboratory of Nuclear Physics and
Technology, \\Peking University, Beijing 100871,
China}\affiliation{\PKU}
\newcommand*{\CHEP}{Center for High Energy
Physics, Peking University, Beijing 100871,
China}\affiliation{\CHEP}
\newcommand*{\CHPS}{Center for History and Philosophy of Science, Peking University, Beijing 100871,
China}\affiliation{\CHEP}


\author{Yuezhen Niu}\affiliation{\PKU}
\author{Junwu Huang}\affiliation{\PKU}
\author{Bo-Qiang Ma}\email{mabq@pku.edu.cn}\affiliation{\PKU}\affiliation{\CHEP}\affiliation{\CHPS}



\begin{abstract}
We study the self-gravitational effect on the equation of
state (EoS) of Bose and Fermi gases in thermal equilibrium at the
end of reheating, the period after quark-hadron transition and
before Big Bang Nucleosynthesis (BBN). After introducing new grand
canonical partition functions based on the work of Uhlenbeck and
Gropper, we notice some interesting features of the newly developed EoSs with
distinct behaviors of relativistic
and non-relativistic gases under self-gravity. The usual negligence
of the self-gravitational effect when solving the background
expansion of the early universe is justified with numerical results,
showing the magnitude of the self-gravitational modification of the
state constant to be less than $O(10^{-78})$. This helps us to
clarify the background thermal evolution of the primordial patch. Such clarification
is crucial in testing gravity theories, evaluating inflation
models and determining element abundances in BBN.
\end{abstract}





\maketitle

\section{Introduction\label{intro}}
In most history of our universe, gravity has been the dominant force
in determining cosmological evolutions. Due to the onset of
gravitational instability, primordial fluctuations generated in
cosmic inflation start growing by attracting dark matter and gases
nearby after the matter radiation equilibrium at a temperature
approximately $4000~\mathrm{K}$~\cite{ScottDodelson2003}.
Consequently, structures of the first generation galaxies are
seeded. Shortly after the forming of proto-galaxies, the hydrogen
and helium gases within them begin to condensate and make the first
stars. In the formation of any astrophysical body, self-gravity is
beyond question one of the key factors in determining the specific
destinies of stars, quasars, black holes, and neutron stars. For
example, the self-gravity of neutron stars is considered using the
affine model~\cite{Carter1995,Wiggins2000}, where the
effective relativistic self-gravitational potential is approximated
with Tolman-Oppenheimer-Volkoff (TOV) stellar structure equations.
Indeed, gravity, though much weaker than the other three fundamental
forces, i.e., strong, weak and electromagnetic forces, with its long
range of action and omnipresent character from equivalence principle
of general relativity, determines most astronomical evolutions. One
persuasive example showing the significance of self-gravity is the
boson star. Though  massive boson stars are prevented from the
Heisenberg uncertainty principle, they can actually be formed in the
early universe when the self-interaction of scalar particles is
considered~\cite{Kaup:1968zz,RemoRuffini69}. Various methods are
subsequently developed to study different features of boson stars.
People deal with a system of self-gravitating bosons either by
solving the Schr\"odinger and Poisson equations in a Newtonian
way~\cite{Ingrosso1988}, or by fully accounting the relativistic
effect using Klein-Gordon and Einstein
equations~\cite{Laine:1998rg}. When the Einstein gravity and the
Klein-Gordon field couple with each other, exotic bosons which
possess a small mass may undergo Bose-Einstein condensation driven
by gravity and collapse to form boson stars in the early universe or
latter in galaxy center as a candidate for dark
matter~\cite{Colpi1986,Jetzer:1991jr,Ho:1999hs}.

If self-gravity can trigger phenomena at the early universe so
dramatic like boson stars or primordial black
holes~\cite{Hawking:1971tu,Carr:1974nx} not mentioned before, the
self-gravitational effect in the primordial patch of the early
universe should not be easily discarded without a robust proof. An
exact evaluation of the self-gravity of relativistic particles at
the extremely high energy scale before Big Bang Nucleosynthesis
(BBN) (not high enough to reach the boson star critical temperature)
becomes unavoidable. However, most discussions of boson stars are at
relatively high energy scale $\sim 1$~GeV~\cite{Bilic:2000ef}, and
at the same time entail the non-zero mass of scalar particles to be
greater than around 10~eV to preserve radial
stability~\cite{Sharma:2008sc}. But during the radiation dominated
era before BBN, main ingredients are massless photons, and the
homogeneity of the early patch is preserved unlike that in boson
stars. As a result, it is difficult to fully perceive the
self-gravitational behavior of photons in the early universe by
simply extending the methods developed for boson stars. Although
there are several attempts to include the self-gravitational effect
of non-relativistic gases at equilibrium, such as Monte Carlo
simulations, analytic mean field methods and low density
expansions~\cite{deVega:1999pj}, no method so far exists to directly
solve the self-gravity of massless particles, i.e., photons. As a
result, we try to develop an appropriate way to deal with the
self-gravity of massless bosons in the early universe. The newly
developed method will be applied to massive particles as well.

Before BBN, the universe can be well described as in thermal
equilibrium with very high temperature and frequent
interactions. Remarkable features of the early universe,
homogeneity, isotropy and thermal equilibrium together can help us
simplify the discussion to a great extent. Since the Equation of
State (EoS) can serve as a good description of a homogeneous
systems, we manage to delve into the problem with statistical
physics. As is well known, once including self-interactions between
particles, the corresponding partition function shall be changed.
Such change further leads to modifications in the EoS. We try to uncover
the self-gravitational effect by examining the corresponding
statistical properties of Bose and Fermi gases first. Whether
corrections from self-gravity are negligible or bear some new
features not considered before shall be tested in detail in this
work.

In the radiation-dominated epoch, gravities between relativistic particles are commonly neglected in
existing models. When the temperature of the universe decreases to
approximately $10^{11}~\mathrm{K}$, particles are in thermal
equilibrium without strong nuclear interactions and the principal
ingredients at that time are photons, neutrinos, electrons and their
anti-particles. With the energy of the universe dominated by
photons, other ingredients, whether massive particles like electrons
and positrons or the massless ones, all behave radiation-like due to
their frequent collisions with photons and each other at the
ultra-high temperature. The matter soup at that time is suggested to
follow the equation of state $P=\rho/3$ without considering
gravitational attractions between particles. From the equivalence
principle, massless particles with both high kinetic energies and
particle densities may possess a considerable gravitational effect. We
deal with self-gravitational interactions between these thermalized
particles in a Newtonian way. With a statistical approach, we can
avoid the theoretical difficulties in solving equations of general
relativity as well as numerical errors in actual realization, and at
the same time, perceive macro-behaviors without any impediment from
micro-complexity. In Sec.~{\ref{Gcp}}, we deduce a new form of grand
canonical function for Bose and Fermi gases in the relativistic and
the non-relativistic limits. In Sec.~{\ref{earlyuni}}, we apply the
newly-developed partition function to the early universe before BBN.
We also analyze the time dependent term of the EoS and set a strict
upper bound on the modifications of the state constant from
self-gravity. Possible influences on relating topics are discussed
in Sec.~{\ref{remark}}.

\section{Grand canonical partition function for self-gravitational Bose and Fermi gases\label{Gcp}}

In standard statistical physics, people deal with interactions
between gases by including $e^{-\phi _{ij}/kT}$ in the grand
canonical partition function, where $\phi _{ij}$ is the potential
energy resulting from the interactions between two particles $i$,
$j$. The potential energies $\phi _{ij}$ are set to be zero in
general. Such approximation works pretty well when interaction is negligible,
that is, the density-temperature ratio is extremely low. On the
other hand, if the density is relatively high like that in the early
universe, the factor from the self-gravitational effect of bosons
and fermions becomes appreciable. As Uhlenbeck and Gropper~\cite{UhlenbeckGropper32} proved,
the partition function of non-ideal Bose or Fermi gases should be
modified as
\begin{equation}
e^{-\phi _{ij}/kT}\left(1\pm \mathrm{exp}\left[-4\pi
mk_{B}Tr_{ij}^{2}/h^{2}\right]\right), \label{eq1}
\end{equation}
where the plus sign is for Fermi gas.

The grand canonical partition function becomes
\begin{widetext}
\begin{equation}
\Xi =\sum_{N}\frac{e^{-\alpha N}}{N!h^{3N}}\prod_{i}\int e^{-\frac{E_{k}}{%
k_{B}T}}\mathrm{d} \mathbf{p}_i\int \prod_{ij}^{i<j}e^{-\frac{\phi
_{ij}}{k_{B}T}}\left(1\pm \mathrm{exp}\left[-4\pi
mk_{B}Tr_{ij}^{2}/h^{2}\right]\right)\mathrm{d} \mathbf{r}_{ij},
\label{eq2}
\end{equation}
\end{widetext}
where $\alpha=-\mu/k_B T$ represents the chemical potential, and $m$ is the
mass of the particle.

\subsection{The non-relativistic case\label{nonrela}}

For non-relativistic particles, which satisfy the dispersion relation
$E_k=p^2/2m$, the first integral in Eq.~(\ref{eq2}) turns
\begin{equation}
\prod_{i}\int e^{-\frac{{p_{i}}^{2}}{2mk_{B}T}}\mathrm{d}
\mathbf{p}_i=\left(2m\pi k_{B}T\right)^{\frac{3N}{2}}. \label{eq3}
\end{equation}
Since in the non-relativistic case,
$\phi_{ij}={Gm_im_j}/{r_{ij}^2}$, the second integral becomes
\begin{widetext}
\begin{equation}
\int \prod_{ij}^{i<j}e^{-\frac{Gm^{2}}{r_{ij}k_{B}T}}\left(1\pm
\mathrm{exp} \left[-4\pi mk_{B}Tr_{ij}^{2}/h^{2}\right]\right)\mathrm{d}
\mathbf{r}_{ij}=\int e^{-\sum_{i<j}\frac{Gm^{2}}{
r_{ij}k_{B}T}}\left(1\pm \sum_{ij} \mathrm{exp}\left[-4\pi
mk_{B}Tr_{ij}^{2}/h^{2}\right]\right)\mathrm{d}\mathbf{r}_{ij}. \label{eq4}
\end{equation}
\end{widetext}

With $\mathrm{exp}\left[-4\pi mk_{B}Tr_{ij}^{2}/h^{2}\right]\ll 1$,
the omissions made in Eq.~(\ref{eq4}) are justified. At the same
time, we use mean field theory to include only the average effect of
$r_{ij}$. The thermal equilibrium used in mean field theory is
guaranteed in the early universe. As $N\rightarrow \infty $ we
approximate $ N(N-1)/2\rightarrow N^{2}/2$, following which
Eq.~(\ref{eq4}) becomes
\begin{eqnarray}
&&\int e^{-\frac{N^{2}Gm^{2}}{2r_{\ast }k_{B}T}}\left(1\pm \frac{N^{2}}{2}\mathrm{exp}
\left[-4\pi mk_{B}Tr_{\ast }^{2}/h^{2}\right]\right)\mathrm{d} \mathbf{r_{\ast }}\mathrm{d} \mathbf{r}  \nonumber \\
&=&V^{N-1}\left\{\int e^{-\frac{N^{2}Gm^{2}}{2r_{\ast
}k_{B}T}}\mathrm{d}\mathbf{r}_{\ast }\pm \int
\frac{N^{2}}{2}\mathrm{exp}\left[-4\pi mk_{B}Tr_{\ast
}^{2}/h^{2}\right]\mathrm{d} \mathbf{
r_{\ast }}\right.  \nonumber \\
&&\left.\pm \int \frac{N^{2}}{2}\left(e^{-\frac{N^{2}Gm^{2}}{2r_{\ast
}k_{B}T}}-1\right) \mathrm{exp}\left[-4\pi mk_{B}Tr_{\ast
}^{2}/h^{2}\right]\mathrm{d} \mathbf{r_{\ast }}\right\}, \label{eq5}
\end{eqnarray}
where $r_{\ast }$ is the average distance between two particles. We
can see that the first term on the right side of Eq.~(\ref{eq5}) represents the
gravitational correction of the grand canonical partition function
in conventional treatment and the second term corresponds to usual
Bose-Einstein or Fermi-Dirac statistics, while the last term is a
newly-introduced coupling term from the gravity of bosons or fermions.
Again, the upper sign is for Fermi gas.

In the early universe, particle number and energy density are
relatively high with temperature$\sim $ $10^{11}~\mathrm{K}$, i.e.,
exp$[-N^{2}Gm^{2}/2 r_{\ast }k_{B}T]\sim 1$. To integrate the first
term of Eq.~(\ref{eq5}), we expand it into the the first-order
Taylor series and get
\begin{eqnarray}
\int e^{-\frac{N^{2}Gm^{2}}{2r_{\ast }k_{B}T}}\mathrm{d}
\mathbf{r_{\ast }} &=&4\pi
\int_{0}^{r_{0}}\left(1-\frac{N^{2}Gm^{2}}{2r_{\ast
}k_{B}T}\right)r_{\ast
}^{2}\mathrm{d}r_{\ast }  \nonumber \\
&=&V-\left(\frac{9\pi}{16}\right)^{1/3}V^{2/3}\frac{N^{2}Gm^{2}}{k_{B}T}.
\label{eq6}
\end{eqnarray}
This reveals that the upper bound of the integration corresponds to
the spacial scale of the system.

The second term of Eq.~(\ref{eq5}) is
\begin{eqnarray}
&&\int \frac{N^{2}}{2}\mathrm{exp}\left[-4\pi mk_{B}Tr_{\ast
}^{2}/h^{2}\right]\mathrm{d} \mathbf{
r_{\ast }}  \nonumber \\
&=&4\pi \int_{0}^{r_{0}} \frac{N^{2}}{2}\mathrm{exp}\left[-4\pi
mk_{B}Tr_{\ast
}^{2}/h^{2}\right]r_{\ast }^{2}\mathrm{d}r_{\ast }  \nonumber \\
&=&4\pi \int_{0}^{\infty} \frac{N^{2}}{2}\mathrm{exp}\left[-4\pi
mk_{B}Tr_{\ast
}^{2}/h^{2}\right]r_{\ast }^{2}\mathrm{d}r_{\ast }  \nonumber \\
&=&2^{-5/2}\pi^{3/2}N^{2}\Lambda_{\mathrm{nonrela}}^{3}, \label{eq7}
\end{eqnarray}
where $\Lambda_{\mathrm{nonrela}}=h/\sqrt{2\pi mk_{B}T}$ is the
thermal de~Broglie wavelength, which should be much smaller than the
radius of our interested system. If the scale of the system is equal
to or smaller than the thermal de~Broglie wavelength, a situation
which might happen at a much earlier time, such approximation may not
be valid, and we should use the error function $\mathrm{erf}(x)$ in the
integration instead. Again applying the first-order Taylor
expansion, we get the remaining term
\begin{widetext}
\begin{equation}
-\frac{N^{4}\pi Gm^{2}}{k_{B}T}\int_{0}^{r_{0}}\mathrm{exp}\left[-4\pi
mk_{B}Tr_{\ast }^{2}/h^{2}\right]r_{\ast }\mathrm{d}r_{\ast }=-\frac{N^{4}
\pi Gm^{2}}{ 4k_{B}T}\Lambda_{\mathrm{nonrela}}^{2}.  \label{eq8}
\end{equation}
\end{widetext}

Finally, we find the new grand canonical partition function
\begin{widetext}
\begin{equation}
\Xi =\left\{1-\frac{Gm^{2}}{k_{B}T}\left(\frac{9\pi
}{16}\right)^{1/3}V^{-1/3}\pm \frac{\pi^{3/2}
\Lambda_{\mathrm{nonrela}}^{3}}{2^{5/2}V}\frac{\partial
^{2}}{\partial \alpha ^{2}}\mp \frac{\pi
Gm^{2}}{4k_{B}TV}\Lambda_{\mathrm{nonrela}}^{2}\frac{\partial
^{4}}{\partial \alpha ^{4}}\right\}\mathrm{exp}\left[e^{-\alpha
}\frac{V}{\Lambda_{\mathrm{nonrela}}^{3}}\right], \label{eq9}
\end{equation}
\end{widetext}
with which we can obtain the total energy by $E=\partial
ln\Xi/\partial \beta$ with $\beta=1/k_{B}T$, and then the entropy.
As suggested above, this partition function is applicable when the
scale of the system is much larger than the thermal de~Broglie
wavelength. However, such premises may be wavered at a much earlier
stage. The first term of the partition function is the grand
canonical function for ideal Boltzmann gas. The second term
represents the self-gravitational correction of Boltzmann gas, and
its magnitude corresponds to the portion that gravitational energy
$E_p\sim Gm^{2}/V^{1/3}$ contributes to the total energy
$E_{\mathrm{total}}\sim k_BT$. In turn it manifests whether
gravitational correction is important in dealing with the evolution
of the universe. The third term is the correction from identical
principle, and its magnitude (proportional to the cubic square of
thermal de~Broglie wavelength over the volume of the system $V$)
evaluates how strong quantum identity is. As a whole, the
self-gravitational effect on fermions is restricted by both the
strength of gravity and the intensity of quantum effect.

\subsection{The relativistic case\label{rela}}

For highly relativistic particles, the rest mass is negligible
compared with the kinetic energy, thus the dispersion
relation becomes $E_k=pc$. Then the grand canonical partition
function is found to be
\begin{widetext}
\begin{equation}
\Xi =\sum_{N}\frac{e^{-\alpha N}}{N!h^{3N}}\int
e^{-\sum_{i}\frac{p_{i}c}{ k_{B}T}}\mathrm{d} \mathbf{p}_i\int
e^{-\sum_{i<j}\frac{Gp_{i}p_{j}}{r_{ij}k_{B}Tc^{2}}}\left(1\pm
\sum_{i<j}\mathrm{exp}\left[-4\pi
^{2}p_{i}k_{B}Tr_{ij}^{2}/h^{2}\right]\right)\mathrm{d} \mathbf{r}_{ij}.
\label{eq10}
\end{equation}
\end{widetext}

We can solve it exactly by diagonalizing
$\sum_{i<j}\frac{Gp_{i}p_{j}}{r_{ij}k_{B}Tc^{2}}$ first. But such
treatment is proved to be unnecessary since, on one hand, the
self-gravitational effect in the relativistic case acts as a
correction term, and on the other hand, for our problem in the
homogeneous and isotropic Robertson-Walker metric background at
exceptionally high temperature, relativistic particles are in a
thermal equilibrium state evolving in a radiation manner resulting
from frequent collisions. Therefore, we expect self-gravity alone
will not significantly destroy this homogeneity directly shown by
CMB observations. All particles in the patch then possess the same
characteristic average momentum due to the frequent collision. We
then approximate the summation as
\[
-\sum_{i}\frac{p_{i}c}{k_{B}T}-\sum_{i<j}\frac{Gp_{i}p_{j}}{r_{ij}c^{2}k_{B}T
}\approx\sum_{i}\left(\frac{c}{k_{B}T}-\sum_{j}\frac{G\bar{p}}{r_{ij}c^{2}k_{B}T}
\right)p_{i},
\]
where $\bar{p}$ is the average momentum of the particles in the
system. So the grand canonical partition function becomes
\begin{widetext}
\begin{eqnarray}
\Xi  &=&\sum_{N}\frac{e^{-\alpha N}}{N!h^{3N}}\{\prod_{i}\int e^{-\left(\frac{c
}{k_{B}T}+\sum_{j}\frac{G\bar{p}}{2r_{ij}c^{2}k_{B}T}
\right)p_{i}}\mathrm{d} \mathbf{p}_{i} \mathrm{d}\mathbf{r}_{i}  \nonumber \\
&&\pm \prod_{i}\int e^{-\left(\frac{c}{k_{B}T}+\sum_{j}\frac{G\bar{p}}{
2r_{ij}c^{2}k_{B}T}+\sum_{j}4\pi
^{2}\frac{k_{B}Tr_{ij}^{2}}{h^{2}c^{2}}
\right)p_{i}}\mathrm{d} \mathbf{p}_{i}\mathrm{d}\mathbf{r}_{i}\}  \nonumber \\
&=&\sum_{N}\frac{e^{-\alpha N}}{N!h^{3N}}\int
\frac{(8\pi)^{N}}{\left(\frac{c}{k_{B}T}
+\sum_{j}\frac{G\bar{p}}{2r_{ij}c^{2}k_{B}T}\right)^{3N}}\mathrm{d} \mathbf{r}_{i}  \nonumber \\
&&\pm \sum_{N}\frac{e^{-\alpha N}}{N!h^{3N}}\int
\frac{(8\pi)^{N}}{\left(\frac{c}{k_{B}T}
+\sum_{j}\frac{G\bar{p}}{2r_{ij}c^{2}k_{B}T}+\sum_{j}4\pi ^{2}\frac{
k_{B}Tr_{ij}^{2}}{h^{2}c^{2}}\right)^{3N}}\mathrm{d}\mathbf{r}_{i}.
\label{eq11}
\end{eqnarray}
\end{widetext}
To integrate the first term in Eq.~(\ref{eq11}), we perform the Taylor
expansion to the first order knowing the magnitude of gravity being
small outside the critical sphere with radius $r_{m}$
\begin{widetext}
\begin{equation}
\int_{r_{m}}^{r_{0}}\frac{(8\pi)^N}{\left(\frac{c}{k_{B}T}+\sum_{j}\frac{G\bar{p}}{
2r_{ij}c^{2}k_{B}T}\right)^{3N}}\mathrm{d} \mathbf{r}_{i}=\int
\frac{(8\pi)^{N}}{\left(\frac{c}{k_{B}T}\right)^{3N}}\left(1-3N\sum_{j}\frac{G
\bar{p}}{2r_{ij}c^{3}}\right)\mathrm{d} \mathbf{r}_{i}.
\label{eq12}
\end{equation}
\end{widetext}
The integration from 0 to $r_{m}=G\bar{p}/c^3$
\begin{equation}
\int_{0}^{r_{m}}\frac{(8\pi)^N}{\left(\frac{c}{k_{B}T}+\sum_{j}\frac{G\bar{p}}{
2r_{ij}c^{2}k_{B}T}\right)^{3N}}\mathrm{d} \mathbf{r}_{i} \label{12+}
\end{equation}
is neglected since its magnitude is proportional to $r_m^{3N+1}$,
much smaller than the first one in our analysis when gravity contributes as
a correction term.

Under the same mean field approximation, Eq.~(\ref{eq12}) becomes
\begin{equation}
\frac{(8\pi)^N}{\left(\frac{c}{k_{B}T}\right)^{3N}}\left(V-\left(\frac{\pi}{6}\right)^{1/3}\frac{9\pi
GN^{2}\bar{p}}{c^{3}} V^{2/3}\right)V^{N-1}. \label{12final}
\end{equation}

Before going on to integrate the second term of Eq.~(\ref{eq11}), we
first compare
the magnitude of different terms in the denominator $\frac{c}{k_{B}T}+\sum_{j}
\frac{G\bar{p}}{2r_{ij}c^{2}k_{B}T}+\sum_{j}4\pi ^{2}\frac{k_{B}Tr_{ij}^{2}}{
h^{2}c^{2}}$. When $\frac{c}{k_{B}T}\backsim 4\pi ^{2}\frac{k_{B}Tr_{ij}^{2}
}{h^{2}c^{2}}$, i.e. $r_{ij}\backsim \frac{hc^{3/2}}{k_{B}T}$, the
corresponding
energy density can be approximated as $U\backsim \frac{hc}{r_{ij}^{4}}%
\backsim 10^{86}$ in the Planck unit. However, the actual energy density
is of magnitude $10^{22} $ ~\cite{Weinbergbook}, meaning that the
first term is $O(10^{-16})$ compared with ${\pi
}^{2}\frac{k_{B}Tr_{ij}^{2}}{h^{2}c^{2}}$. Besides, since in our
interested time period, the temperature varies within one order in
magnitude, so long as the number density does not change beyond
$O(10^{21})$, i.e. the volume of the universe increases by less than
$O(10^{21})$ within 1s (such circumstances are fully
guaranteed in a classical radiation-dominated universe), $
\frac{c}{k_{B}T}$ can then be safely neglected. But for some earlier
stages this treatment may fail. In the same way,
$\frac{G\bar{p}}{r_{ij}c^{2}k_{B}T}$ is $O(10^{-48})$ compared with
the third term and can be justifiably omitted if the number density
varies within $O(10^{40})$. Neglecting the first two terms after
careful magnitude evaluation, we have
\begin{widetext}
\begin{equation}
\int
\frac{(8\pi)^N}{\left(\frac{c}{k_{B}T}+\sum_{j}\frac{G\bar{p}}{r_{ij}c^{2}k_{B}T}
+\sum_{j}4\pi
^{2}\frac{k_{B}Tr_{ij}^{2}}{h^{2}c^{2}}\right)^{3N}}\mathrm{d}\mathbf{r}_{i}\ll
\frac{(8\pi)^N}{\left(\frac{c}{k_{B}T}\right)^{3N}}\left(V-\left(\frac{\pi}{6}\right)^{1/3}\frac{9\pi
GN^{2}\bar{p}}{c^{3}} V^{2/3}\right)V^{N-1}.  \label{eq13}
\end{equation}
\end{widetext}

Therefore only the first integration contributes to our final
result, corresponding to a unification of Bose and Fermi
distributions under the relativistic limit. The grand canonical
partition function for relativistic particles including the
self-gravitational effect becomes
\begin{widetext}
\begin{equation}
\Xi =\left[1-\left(\frac{\pi }{6}\right)^{1/3}\frac{9\pi
Gk_{B}T}{c^4V^{1/3}}\frac{\partial ^{2}}{\partial \alpha
^{2}}\right]\mathrm{exp}\left[e^{-\alpha }{\left(\frac{8\pi
k_{B}T}{ch}\right)}^{3}V\right] =\left[1-6^{2/3}\frac{3\pi
Gh}{\Lambda_\mathrm{rela}c^3V^{1/3}}\frac{\partial ^{2}}{\partial
\alpha ^{2}}\right]\mathrm{exp}\left[e^{-\alpha
}\frac{V}{(\Lambda_\mathrm{rela})^3}\right], \label{eq14}
\end{equation}
\end{widetext}
where $\Lambda_\mathrm{rela}=ch/2{\pi}^{1/3}k_BT$ is the thermal
de~Broglie wavelength for massless particles. In Eq.~(\ref{eq14}),
the first term is merely the ordinary ideal Boltzmann gas term,
while the second is induced by self-gravity, proportional to
$Gh/\Lambda_\mathrm{rela}c^3V^{1/3}$, i.e. the gravitational
potential energy over the self-energy of particles. By comparing the
two terms we can catch a glimpse of how gravity acts on the
statistical properties of the relativistic particles.

\section{Result in the early universe\label{earlyuni}}

Before applying the newly developed method to a specific background,
we first review the whole evolution of the early universe. In the
very early stages of the evolution, to resolve several problems in
the Big Bang cosmology, like the horizon problem, magnetic-monopole
problem, structure formation problem and so on, the universe is
suggested to undergo an inflation between time scales
$10^{-43}~\mathrm{s}$ and $10^{-34}~\mathrm{s}$. By the end of the
inflation, the background temperature is roughly
$10^{28}~\mathrm{K}$ when strong, weak and electromagnetic forces
are unified. As the temperature cools down, the strong force freezes out,
weak and electromagnetic interactions are subsequently off stage and
before $10^{-4}~\mathrm{s}$. Afterward, neutrinos decouple at
$T\sim10^9~\mathrm{K}$. Then Big Bang Nucleosynthesis begins at
temperature $10^9~\mathrm{K}\to 10^3~\mathrm{K}$. At the end of BBN,
photons decouple from matter, and form the cosmic microwave
background radiation we observe today. To use our statistical method
in studying the self-gravitational effect, assumptions of
homogeneity, isotropy and thermal equilibrium should be justified.
We thus choose to study the universe before BBN when three criteria
are largely met and the background temperature $T_{\mathrm{th}}$ is
of magnitude $10^{11}~\mathrm{K}$ ~\cite{Weinbergbook}, well below
the corresponding thermal energy of the rest mass of $\pi$ mesons
and strong interactions can be neglected. Matter and radiation in
the universe at that time are in thermal equilibrium, resulting from
the rapid collisions in the exceedingly high temperature despite of
the fast universal expansion. By identifying particles whose rest
masses are below the corresponding thermal energy of the background
temperature $T_{\mathrm{th}}$, we find the main ingredients to be
electrons, neutrinos, their anti-particles and photons. These
abundant genres are highly relativistic. For example, the rest
energy of the electron corresponds to the thermal energy at
temperature $\sim10^9~\mathrm{K}$, two orders of magnitude smaller
than the background temperature. And non-relativistic particles,
mainly protons and neutrons, constitute only an extremely small
portion of the whole ingredients, approximately $10^{-10}$ as many
as that of relativistic ones. Below we deal with relativistic and
non-relativistic particles separately, though the latter is unable
to significantly influence the background evolution.

\subsection{Thermalized relativistic particles\label{Threla}}

At a temperature higher than $8000~\mathrm{K}$, the universe is
radiation dominated. Consequently, relativistic particles are
undoubtedly the main ingredients during our studying period with a
temperature approximately $10^{11}~\mathrm{K}$. And both massive and
massless components act in a radiationlike manner due to frequent
collisions. We use the Planck unit $k_B=h=c=G=1$ in all calculations
and some of our discussions below. Moreover, the volume
$V_\mathrm{init}$, which is about four light years in radius in our
partition function, corresponds to the biggest thermalized region
constrained by CMB anisotropy and grows as the extension of particle
horizon of the primordial patch. Though the exact size of the early
universe at this time is not determined yet. The time span of our
study is $0.1~\mathrm{s}\sim 1~\mathrm{s}$. We consider the
relativistic bosons first and later find that fermions behave
similarly as expected. Their grand canonical partition function is
\begin{equation}
\Xi =\left[1-6^{2/3}\frac{3\pi
Gh}{\Lambda_\mathrm{rela}c^3V^{1/3}}\frac{\partial ^{2}}{\partial
\alpha ^{2}}\right]\mathrm{exp}\left[e^{-\alpha
}\left(\frac{V}{(\Lambda_\mathrm{rela})^3}\right)\right],
\label{eq25}
\end{equation}
The chemical potential is set to be zero since the particle number
of the photon, which is the main ingredient, does not conserve. Here,
instead of directly solving the Friedman equation with the density
and pressure derived in the new grand canonical partition function,
we approximate the temperature evolution the same as that in
the classical radiation-dominated model neglecting the
self-gravitational effect
\begin{equation}
T=at^{-1/2},  \label{eq26}
\end{equation}
where $a$ is a constant to be determined by the initial condition.
Combining Eqs.~(\ref{eq25}) and (\ref{eq26}), we numerically find
that the density of relativistic particles relies on time as
\begin{equation}
\rho =\frac{b_{1}}{t^2}+ \frac{\tilde{b}_1}{\sqrt{t} V},
\label{eq27}
\end{equation}
where $b_{1}$ and $\tilde{b}_1$ are separately
$6.58*10^{150}~\mathrm{J\cdot s^2/m^3}$ and
$7.09*10^{18}~\mathrm{J\cdot s^2}$. This differs from the
conventional result without considering the self-gravitational
effect where $\rho \sim T^{4}\sim 1/t^{2}$. That is, once including
the self-gravity, energy density decreases much slower. This
disparity in energy density can be explained as an increase in
gravitational potential caused by the rapid elongation of particle
separation in cosmic expansion. This remedies the conventional
energy lost with the decrease of particle density and the extension of
particle wavelength. The pressure behaves as
\begin{equation}
P =\frac{b_{1}}{3t^2}+\frac{\tilde{b}_2}{\sqrt{t} V}, \label{eq27+}
\end{equation}
where $\tilde{b}_2=1.67*10^{18}~\mathrm{J\cdot s^2}$. As a result,
the state constant no longer stays invariant but instead relies on
time
\begin{equation}
w =\frac{1}{3}\left(1-\frac{d_1}{\sqrt{t}V}\right), \label{eq27++}
\end{equation}
where $d_1$ has a numerical value $3.97\times10^{30}$ in Planck unit
which is $9.69\times10^{-95}~\mathrm{s}^{1/2}\cdot \mathrm{m}^3$.
Knowing that during our studying period $t\sim10^{42}$ in Planck
unit, or $0.1s$ in standard unit, and the volume is $V\sim10^{156}$
in Planck unit, or corresponds to four light years in radius
justified by CMB observation, we find that the correction term in
Eq.~(\ref{eq27++}) is of the magnitude $O(10^{-147})$. Since the
universe is radiation dominated at this time, we find the correction
term $d_1/\sqrt{t}V\sim t^{-1}$. Differing from the non-relativistic
case below, the modification from self-gravity rapidly diminishes as
time goes on and volume expands.

Besides, we can roughly estimate the modification term at much
earlier times, such as the end of inflation when $t\sim
10^{-34}~\mathrm{s}$ and $V\sim 10^{-51}V_\mathrm{init}$, to be
$O(10^{-79})$. Thus the estimated magnitude of the modification is
comparable to the effect in the non-relativistic case.

\subsection{Thermalized non-relativistic nuclear particles\label{Tnonrela}}

Although  massive nuclear particles contribute only a negligible
portion to the whole energy makeup during the period we study, their
abundances are crucial in determining the late time nuclear
synthesis. As a result of the rapid expansion, the energy density of
the radiation decreases swiftly as $a^{-4}$ in classical treatment,
while matter density such as that of protons and neutrons decreases
much slower as $a^{-3}$. Additionally, relativistic particles and
their anti-particles annihilate with each other as temperature goes
down, and the non-relativistic nuclear particle component, therefore,
becomes increasingly important. The non-relativistic feature of
nuclear particles shall preserve as temperature decreases throughout
the thermalization and into nuclear synthesis. Since most of the
nuclear particles are fermions, we consider only non-relativistic
fermion gases without loss of generality. We can safely neglect the
chemical potential of Bose gases since the photon number is not
conserved. But for non-relativistic Fermi gases, at extremely high
temperature, such a simplification may no longer be appropriate.
Fortunately, it turns out that the chemical potential fails to show
up in density and pressure expressions to the first and second order
approximations in our result. This not only significantly simplifies
the calculation, but also justifies the usual negligence of chemical
potential in the evolution of non-relativistic fermions in the early
universe~\cite{ScottDodelson2003}.

For massive fermions, the grand canonical partition function is
\begin{widetext}
\begin{equation}
\Xi =\left\{1-\frac{Gm^{2}}{k_{B}T}\left(\frac{9\pi
}{16}\right)^{1/3}V^{-1/3}\pm \frac{\pi^{3/2}
\Lambda_{\mathrm{nonrela}}^{3}}{2^{5/2}V}\frac{\partial
^{2}}{\partial \alpha ^{2}}\mp \frac{\pi
Gm^{2}}{4k_{B}TV}\Lambda_{\mathrm{\mathrm{nonrela}}}^{2}\frac{\partial
^{4}}{\partial \alpha ^{4}}\right\}\mathrm{exp}\left[e^{-\alpha
}\frac{V}{\Lambda_\mathrm{nonrela}^{3}}\right]. \label{eq17}
\end{equation}
\end{widetext}
We then obtain the energy density in a conventional way
\begin{equation}
U=-\frac{\partial }{\partial \beta }ln\Xi,   \label{eq18}
\end{equation}
\begin{equation}
\rho =-\frac{U}{V}=\rho (\beta ,V).  \label{eq19}
\end{equation}
That is, the density is a function of $\beta $ and volume, while
$\beta$ and V vary in time. Here the background temperature
evolution can be well approximated by $T\sim t^{-1/2}$ during the
radiation dominated period. In the time span we are studying, $\beta
$ varies within one order of magnitude. To simplify the problem
correspondingly, we set the temperature to be $10^{11}~\mathrm{K}$
and solve the density evolution. The mass of general
non-relativistic particles in calculation is chosen to be the
neutron mass, since masses of other non-relativistic nuclear
particles at that time differ only by a coefficient within one order
of magnitude.

\begin{figure}[t]
\includegraphics[width=0.5\textwidth]{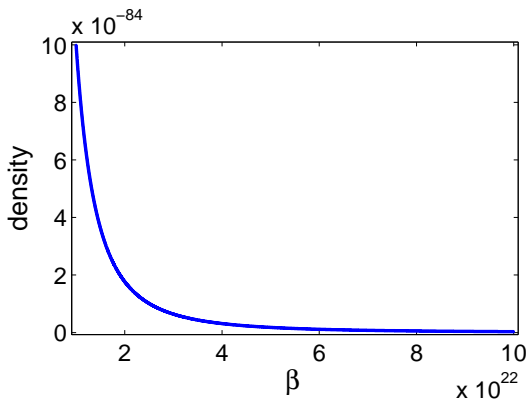}
\caption{The reduced density $\beta$ relation in Planck unit.}
\label{fig:dis massive}
\end{figure}

\begin{figure}[th]
\centering
\includegraphics[width =\columnwidth]{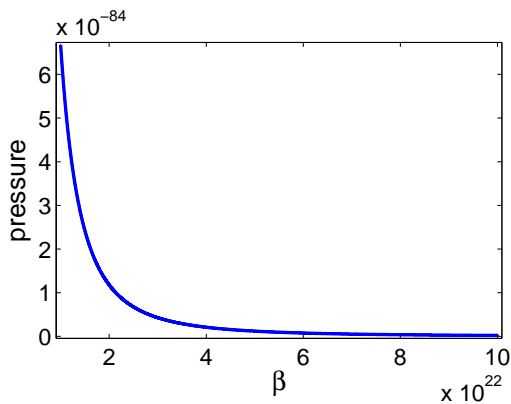}
\caption{The reduced pressure $\beta$ relation in Planck unit.}
\label{fig:dis pressure}
\end{figure}

The expression for $\rho$ becomes exceedingly lengthy. Facing a huge
magnitude range which unavoidably exceeds the calculation limit in the
computer, to evaluate the expression without any numerical reduction
is impossible. However, after observing the density behavior within
a small volume range, an asymptotic behavior is found. We then
analyze the density behavior as volume radius increases to four light
years. After careful magnitude analysis, we find that as long as the
volume of the system exceeds $10^{-70}V_\mathrm{init}$, where
$V_\mathrm{init}$ is the initial volume of the primordial patch.
Such a requirement is thus safely guaranteed in the primordial patch
before BBN. The first-order density behavior relying on the
temperature is found to be
\begin{equation}
\rho=c_1\left(\frac{1}{\beta }\right)^{5/2}=c_1T^{5/2}, \label{eq20}
\end{equation}
where $c_1$ equals $1.27\times10^{-28}$ in Planck unit, or
$3.9\times 10^3~\mathrm{J\cdot K^{5/2}/m^3}$. We use the extreme
condition of the early universe between 0.1~s and 1~s to numerically
reduce the expression and find that the result goes back to the
ideal fermion case, as is shown in Fig.~\ref{fig:dis massive}.
Applying the same method, a similar asymptotic behavior of the first
order can be found in the pressure-volume relation. We notice that
strange negative pressure occurs in the transplanckian scale due to the
failure of statistical physics. However, as the radius of the patch grows to four
light years, which is
the initial value of our calculation, the pressure again becomes almost
volume-independent. The reduced pressure is
\begin{equation}
P=c_2\left(\frac{1}{\beta }\right)^{5/2}=c_2T^{5/2}, \label{eq21}
\end{equation}
where $c_2=2.60\times 10^3~\mathrm{J\cdot K^\frac{5}{2}/m^3}$, as is
shown in Fig.~\ref{fig:dis pressure}.

Then the equation of state in the specific time span becomes
\begin{equation}
\rho =wP,  \label{eq22}
\end{equation}
where the state constant $w=1.5 $ to the first-order approximation,
as can be seen in Fig.~\ref{fig:dis massive}. The recurrence back to
classical case implies that the gravitational effect on
non-relativistic gases is quite small and can be safely neglected to
the first order. We carry on the calculation to the second order and
find the density and the pressure behave as
\begin{equation}
\rho=c_1\left(\frac{1}{\beta }\right)^{5/2}+\frac{\tilde{c}_1}{\beta},
\label{eqden2}
\end{equation}
\begin{equation}
P=c_2\left(\frac{1}{\beta }\right)^{5/2}-\frac{\tilde{c}_2}{\beta},
\label{eqpre2}
\end{equation}
where $\tilde{c}_1$ and $\tilde{c}_2$ have numerical values
$1.37*10^{-70}~\mathrm{J\cdot K/m^3}$ and
$2.06*10^{-77}~\mathrm{J\cdot K/m^3}$ respectively. The state
constant to the second order evolves in time
\begin{eqnarray}
w=\frac{2}{3}\left(1-\frac{c_3}{T^{3/2}}\right),\label{w1}
\end{eqnarray}
where $c_3$ equals $3.54*10^{-25}~\mathrm{K^{3/2}}$, with
temperature $T\sim10^{11}K$. Thus we find the total modification
induced by the self-gravitational effect in the EoS to the second
order acts as an additional term of the magnitude $O(10^{-89})$ and
increases over time as $c_3/T^{3/2}\sim t^{3/4}$. Though relatively
small, we can see that such a term tends to decrease the pressure
between non-relativistic particles as a manifestation of
gravitational attraction. Besides, the deviation grows as
temperature cools down, and finally regains its power in the matter-dominated
era. The whole analysis serves as a robust test to the
justification of discarding self-gravity when solving the evolution
of the non-relativistic nuclear particles during the radiation-dominated period.

\section{Remarks\label{remark}}
In general relativity, any kind of dynamic evolution, if inducing
local energy fluctuation, cannot escape the gravitational
back-reaction, i.e. the change of space time geometry itself. From
the principle of minimal coupling, such dramatic feature of general
relativity indeed does not affect different kinds of interactions
significantly enough in most cases. Consequently, the theory of
general relativity is poorly tested except in a weak-field limit
within small regions like solar systems and radio pulsars, no bigger
than galaxy scale. However, the Friedman equation using
Robertson-Walker metric from general relativity is widely utilized
in solving the homogeneous and isotropic expansion of the universe.
A test of general relativity in a large scale of cosmic size then
proves crucial and necessary in order to justify the foundation of
modern cosmology. One of some promising methods is to use particle
element abundances~\cite{Daniel:2008et} to restrict expansion rate
before BBN.

The weak interaction before BBN between neutrons and protons should
be frequent enough to counter expansion in maintaining equilibrium.
Accordingly, once the expansion speeds up, the weak interaction
between neutrons and protons will be out of equilibrium at higher
temperature, causing a larger neutron-to-proton ratio. In turn, the
abundances of deuterium, helium-3, and helium-4 are increased.
Knowing that helium-4 is hard to destroy during galaxy formation,
observations of helium-4 abundance can help test the expansion rate
and the cosmological constant as previous work
suggested~\cite{Daniel:2008et}. Modifications of the Friedman
equation in their work can be parameterized as
\begin{equation}
\left(\frac{\ddot{a}}{a}\right)=-\frac{4 \pi G}{3}\rho\left[1+\chi\left(
P/3\rho c^2\right)\right],
 \label{modifiedFR}
\end{equation}
where $\chi$ corresponds to the modified state constant $w$ in our
case, see Eq.~(\ref{eq27++}) and Eq.~(\ref{w1}). In existing
observation results, such as those by WMAP~\cite{Spergel:2006hy}, Peimbert {\it et~al.}~\cite{Peimbert:2007vm}, and Steigman~\cite{Steigman:2007xt}, the accuracy of the constraint of
$\chi$ is of the magnitude $O(0.1)$, well beyond the self-gravitational
effect in our result. Thus, we can exclude self-gravity as the
dominant mechanism behind any observable deviation in this method.

Finally, reexamining the derived formula Eq.~(\ref{eq27++}) for
state constant, we compare it with the result in boson
stars~\cite{Colpi1986}, and find that $w$ in both cases increases as
the volume of the system expands. The merging similarities of the
self-gravitational effect between two distinct circumstances are
illuminating to understand the fundamental properties of gravity
itself.

\section{Conclusion\label{conclusion}}

In this paper, we try to evaluate the self-gravitational effect in
Bose and Fermi gases at the early universe and set an exact bound on
the magnitude of the modification. We study the Bose and Fermi gases
before BBN synthesis in thermal equilibrium at a temperature in the
vicinity of $10^{11}~\mathrm{K}$, when particles collide
relativistically with each other frequently enough to maintain
thermalization in spite of the significant cosmic expansion. We
formulate new grand canonical partition functions based on the proof
Uhlenbeck and Gropper made on non-ideal Bose and Fermi gases. Taking
advantage of the severe state of the early universe, we reduce the
modified density and pressure expressions and derive the equations
of state of relativistic and non-relativistic gases, discovering
that self-gravity manifests itself in a simple way. In both
non-relativistic and relativistic cases, we find the deviation from
ideal ones to be small. This justifies the neglecting of the
self-gravitational effect when dealing with non-relativistic
particles. However, dynamical behavior appears when considering the
second order effect. Our result serves as a robust proof for
neglecting self-gravity when solving the evolutions of relativistic
Bose and Fermi gases as well as the thermalized non-relativistic
Fermi gases. Additionally, we distinguish different effects of
self-gravity on the relativistic and the non-relativistic particles.
In thermalized relativistic particles, discrepancies between boson
and fermion statistics become unapparent, and the self-gravitational
effect of magnitude $O(10^{-147})$ decreases with time. This is
naturally expected as matter becomes more and more dominant in the
energy content while the energy density of radiation undergoes a
more rapid dilution. Gravitational attraction consequently
diminishes much faster in radiation components than in matter. While
in non-relativistic Fermi gases, the self-gravitational effect,
though still very small of the magnitude $O(10^{-89})$, increases
over time as the temperature cooling down dramatically and finally
becomes significant in the matter-dominated era when both homogeneity
and isotropy break down and the EoS description is no longer
appropriate. As to the evolution of the background metric, we find
that self-gravity of any kind will not strongly affect the
conventional result using EoS of ideal relativistic and
non-relativistic boson and fermion statistics. Thus the previously
not fully justified negligence of self-gravity is theoretically
studied from a statistical perspective when dealing with cosmic
evolution before BBN. However, our study is restricted to a given
temperature range, and the same result may no longer be true at the
very early stage of the universe as can be seen in
Eq.~(\ref{eq27++}). At a much earlier epoch, when the volume of the
universe is small enough with the basic assumption of our
statistical approach of thermal equilibrium still satisfied,
self-gravity may no longer be negligible from our result. Thus,
further study should be carried out at much earlier stages to give a
complete evaluation on the effects of self-gravity and then the
back-reaction.

\begin{acknowledgments}
We thank Pisin Chen for his comments on the issue at the early
stages of this work. Y.N. acknowledges enlightening discussions with
Alexei Starobinsky. This work is supported by the National Natural
Science Foundation of China (Grant Nos. 11021092, 10975003, and
11035003), and the National Fund for Fostering Talents of Basic Science
(Grant Nos. J0630311, J0730316). It is also supported by the Education
Foundation for Undergraduate Research and Hui-Chun Chin and
Tsung-Dao Lee Chinese Undergraduate Research Endowment (Chun-Tsung
Endowment) at Peking University.

\end{acknowledgments}


\begin{thebibliography}{00}

\bibitem{ScottDodelson2003}
  S.~Dodelson, {\it Modern Cosmology}
  (Academic Press, San Diego, 2003).

\bibitem{Carter1995}
B.~Carter and J.P.~ Luminet, Mon.\ Not.\ R.\ Astron.\ Soc.
{\bf 212}, 23 (1985).

\bibitem{Wiggins2000}
  P.~Wiggins and D.~Lai, Astrophys. J. {\bf 532}, 530 (2000)
  [arXiv:astro-ph/9907365].

\bibitem{Kaup:1968zz}
  D.~J.~Kaup,
  Phys.\ Rev.\  {\bf 172}, 1331 (1968).

\bibitem{RemoRuffini69}
R.~Ruffini and S.~Bonazzola,
  Phys.\ Rev.\  {\bf 187}, 1767 (1969).


\bibitem{Ingrosso1988}
G.~Ingrosso and R.~Ruffini, Nuovo\ Cimento\ B {\bf 101}, 369 (1988).

\bibitem{Laine:1998rg}
  M.~Laine and M.~E.~Shaposhnikov,
  Nucl.\ Phys.\  B {\bf 532}, 376 (1998)
  [arXiv:hep-ph/9804237].

\bibitem{Colpi1986}
M.~Colpi, S.~L.~Shapiro, and I.~Wasserman, Phys.\ Rev.\ Lett. {\bf
57}, 2485 (1986).

\bibitem{Jetzer:1991jr}
  P.~Jetzer,
  Phys.\ Rept.\  {\bf 220}, 163 (1992).

\bibitem{Ho:1999hs}
  J.~Ho, S.~j.~Kim, and B.~H.~Lee,
  [arXiv:gr-qc/9902040].

\bibitem{Hawking:1971tu}
  S.~W.~Hawking,
  Phys.\ Rev.\ Lett.\  {\bf 26}, 1344 (1971).

\bibitem{Carr:1974nx}
  B.~J.~Carr and S.~W.~Hawking,
  Mon.\ Not.\ Roy.\ Astron.\ Soc.\  {\bf 168}, 399 (1974).

\bibitem{Bilic:2000ef}
  N.~Bilic and H.~Nikolic,
  Nucl.\ Phys.\  B {\bf 590}, 575 (2000)
  [arXiv:gr-qc/0006065].

\bibitem{Sharma:2008sc}
  R.~Sharma, S.~Karmakar, and S.~Mukherjee,
  [arXiv:gr-qc/0812.3470].

\bibitem{deVega:1999pj}
  For a review, see H.~J.~de Vega, and N.~G.~Sanchez,
  Phys.\ Lett.\  B {\bf 490}, 180 (2000)
  [arXiv:hep-th/9903236].

\bibitem{UhlenbeckGropper32}
  G.~E.~Uhlenbeck and L.~Gropper,
  Phys.\ Rev.\  {\bf 41}, 79 (1932).

\bibitem{Weinbergbook}
S.~Weinberg, {\it The first three minutes: a modern view of the
origin of the universe} (Basic Books, New York, 1993).

\bibitem{Daniel:2008et}
  S.~F.~Daniel, R.~R.~Caldwell, A.~Cooray, and A.~Melchiorri,
  Phys.\ Rev.\  D {\bf 77}, 103513 (2008)
  [arXiv:astro-ph/0802.1068].

\bibitem{Spergel:2006hy}
  D.~N.~Spergel {\it et al.}  [WMAP Collaboration],
  Astrophys.\ J.\ Suppl.\  {\bf 170}, 377 (2007)
  [arXiv:astro-ph/0603449].

\bibitem{Peimbert:2007vm}
  M.~Peimbert, V.~Luridiana, and A.~Peimbert,
  Astrophys.\ J.\  {\bf 666}, 636 (2007)
  [arXiv:astro-ph/0701580].

\bibitem{Steigman:2007xt}
  G.~Steigman,
  Ann.\ Rev.\ Nucl.\ Part.\ Sci.\  {\bf 57}, 463 (2007)
  [arXiv:astro-ph/0712.1100].




\end{thebibliography}
\end{document}